# The hidden side of paintbrushes: a geometry shaped by power laws

Raphaelle Taub[1], Jérôme Crassous[2,3], Thomas Salez[4,5], Frédéric Restagno[1], Christophe Poulard[1]*

**Affiliations:**

[1]Université Paris-Saclay, CNRS, Laboratoire de Physique des Solides ; F-91405, Orsay, France.

[2]Univ Rennes, CNRS, IPR (Institut de Physique de Rennes)- UMR 6251 ; F-35000 Rennes, France.

[3]PMMH, CNRS, ESPCI Paris, Université PSL, Sorbonne Université, Université de Paris, Paris, 75005, France.

[4]Univ Bordeaux, CNRS, LOMA, UMR 5798 ; F-33405 Talence, France.
[5]Mechanics Department, Ecole Polytechnique, Institut Polytechnique de Paris, 91128 Palaiseau, France.

*Corresponding author. Email: christophe.poulard@universite-paris-saclay.fr.

**Abstract:**
Since prehistoric times, painters have used brush-like tools made from animal and plant materials to apply pigments to rock surfaces. Over time, techniques for creating brushes evolved, and converged towards optimized shapes. Our study performed over hundreds of brushes reveals universal scaling laws. In particular, we uncover a relation between bristle length and diameter linked to the shape of the bristle ends, which governs their mechanical response with and without paint. For round brushes, we found that the size of the paint stain scales with the compression distance, meaning it only depends on the force applied by the painter and not on the mechanical responses of the brushes. These results shed light on the underlying universal physical principles governing brush design and their impact on artistic techniques.

## Main text

The search for universal principles governing the shape of tools is not new. More than a century ago, D'Arcy Wentworth Thompson proposed in *On Growth and Form* that the coarse-grained behavior of both living and artificial systems might obey simple, quantifiable laws, independent of their specific material composition or function (*1*). This visionary perspective has since inspired countless studies across disciplines, revealing that geometry and mechanics often converge toward optimal solutions under physical constraints.

In biology, allometric scaling laws exemplify this principle. Kleiber's law, for instance, demonstrates that the basal metabolic rate of mammals and birds scales with their mass $M$ over many orders of magnitude, illustrating how energetic and geometric constraints shape form and function (*2–4*). Similarly, in biomechanics, Gazzola et al. showed that the scaled swimming speed of aquatic swimmers—from millimetric larvae to blue whales—follows two simple scaling laws when plotted against a scaled swimming frequency, corresponding to a transition from laminar to turbulent locomotion (*5*). These relations provide a compact, physical description of a large diversity of organisms.

Such universal principles extend to technical objects, whose geometries are progressively optimized through empirical evolution. Classic examples include the scaling of nails or turbine blades, where length, diameter, and cross-sectional shape follow robust power-law relations as size increases, reflecting constraints of strength, stiffness, and/or manufacturability (*6*). Similarly, the cables of suspension bridges scale in cross-sectional area with span length to maintain constant tensile stress (*7*), while the wings of flying animals and aircraft converge toward elongated, tapered shapes to optimize lift-to-drag ratios (*8*). Even wind turbine blades follow power-law scaling in chord length and twist distribution to maximize aerodynamic efficiency across sizes (*9*).

Thus, many works aim to describe the shape adopted by bundles of numerous slender beams. This question arises, for example, in the modeling of hair, whether to explain the shape of attached hairs—which depends on the natural curvature of individual strands (*10*)—or to simulate artificial hairstyles for the animation industry (*11–13*). The mechanical and frictional properties of aligned fiber structures, such as those found in pile fabrics like velvet or fleece, also inform the performance and tactile properties of textiles, which are critical for the textile industry (*14*, *15*). In such macroscopic systems, the order and individual shape variations of the fibers often play a crucial role. This is particularly evident in the elastic response of fiber bundles compressed perpendicular to their main axis (*16*).

Paintbrushes, despite their apparent simplicity, are no exception. The emergence of robotic painting and calligraphy has underscored the need for a deeper understanding of paintbrush mechanics. Large-scale analyses of artistic styles have revealed statistical patterns in brushstrokes (*17*), while digital tools have expanded creative possibilities (*18*). Although Herczyński et al. showed how applicator motion, paint properties, and hydrodynamic instabilities shape the traces produced by drops and viscous jets (*19*), the coupled interaction between brush bristles, paint, and the substrate remains poorly characterized. Existing robotic systems rely on empirical calibration to mimic human-like strokes (*20*), and even recent physically motivated models link stroke width to applied pressure without a complete description of brush geometry or rheology (*21*).

These observations motivate a closer examination of paintbrushes, which consist of a handle, a ferrule, and bristles. If one imagines a paintbrush that is as conventional as possible, it is likely to have a spindle shape with a fine tip and an aspect ratio that is difficult to characterize precisely, yet instinctively recognizable. This brush will then probably resemble that of a standard range of round brushes, such as the one pictured in Figure 1. These standard brushes are usually referred to by manufacturers simply as round brushes, and they are ideal for making small strokes, dots, outlines, and sinuous shapes. This apparent convergence toward a standard form raises the question of whether universal physical principles, akin to those seen in biological and engineered systems, might also govern brush design. In addition, the ferrule can be either circular or flattened, taking on a shape resembling a rectangle with more or less rounded corners.

The bristle is the most variable part of a paintbrush, as it directly determines the mechanical and aesthetic properties of the tool. Several raw materials are used. Some are synthetic fibers made from different types of polymers, such as polyamides, polyethylenes, or polypropylenes. Each manufacturer develops its own synthetic fibers, whose composition is often unknown and protected. Natural fibers, sourced from animals like pigs, horses, or mustelids (e.g., Kolinsky red martens), exhibit inherent curvature and surface roughness that influence their bending and frictional behavior. These material-dependent properties, combined with the geometric arrangement of bristles, govern the overall mechanical response of the brush. Whether natural or synthetic, the fibers are typically aligned and bundled in a manner reminiscent of other fiber-based systems, such as hair or textile piles, where collective mechanical properties emerge from individual fiber interactions

**Paintbrush proportions**

In practice, fine art supply stores offer a variety of brushes, both round or flat, with diameters ranging from approximately 1 mm to 25 mm and lengths from about 3 mm to 50 mm. These brushes are not always perfectly round and are often categorize by their specific uses, such as rigger, tracer or retouching brushes or, more specialized types like repique, petticoat, fan, filbert and lettering brushes.

Despite these distinctions in size, fiber material, and naming conventions, the overall proportions of brushes appear remarkably consistent. To investigate whether a universal shape might emerge across these varieties, we analyze the aspect ratios of different brushes in detail in Figure 1. Our study includes fine-art brushes, make-up brushes, nail art brushes, and construction brushes, focusing exclusively on those intended for painting—that is, for applying a liquid onto a surface.

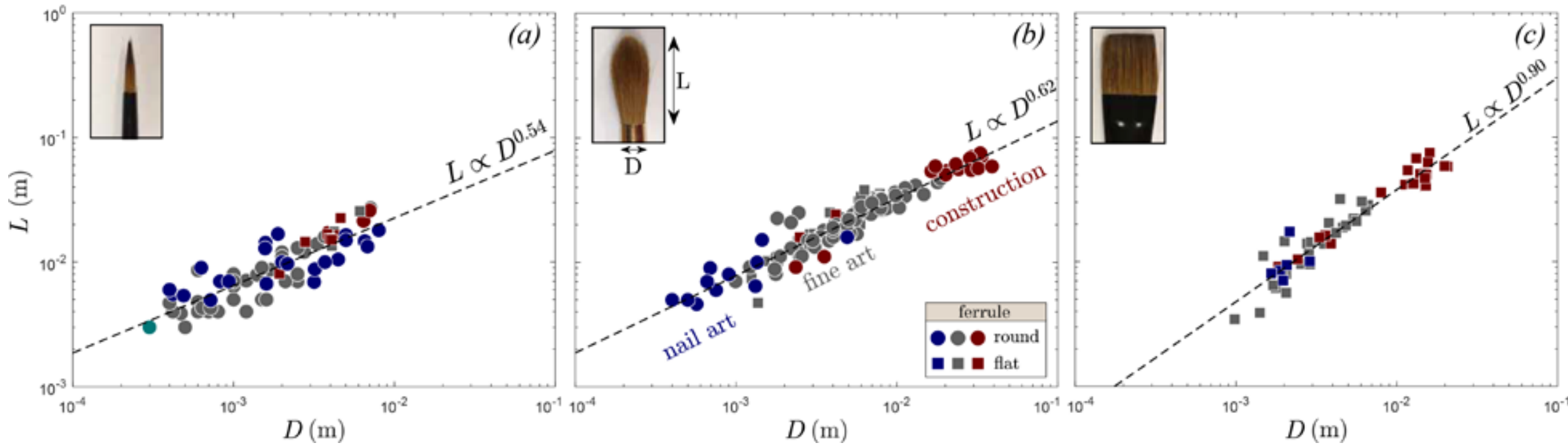

*Figure 1:* ***Length of a random selection of brushes as a function of the ferrule diameter for different intended uses (nail art, fine art, or construction) and tip shapes****. Each point corresponds to one commercial brush. The symbol shape encodes the ferrule geometry (● for round ferrules, ■ for flat ferrules), while the color distinguishes the intended use (nail art, fine art, or construction). The brushes exhibit three main categories of tip geometries: conical ends (a), rounded ends (b), and flattened ends (c). The dashed lines indicate power-law fits for Equation (1) for each tip category, with fitted exponents $\beta_c \simeq 0.54$, $\beta_r \simeq 0.62$, and $\beta_f \simeq 0.90$. The insets show a typical fine-art brushes illustrating the definition of the length L and the ferrule diameter D.*

The diameters and lengths of brushes are measured across wide ranges, with diameters spanning two decades, from 0.4 mm to 40 mm. The length $L$ is define as the distance between the ferrule and the tip of the brush. The diameter $D$ is measured at the ferrule for round brushes, while for flat brushes, it correspond to the width of the ferrule, specifically the shortest dimension of its rectangular cross-section.

The selection of brushes for measurement is designed to include various manufacturers, paint styles, and trim materials (synthetic, natural, and hybrid). In total, approximately 200 brushes are measured from various manufacturers. The collected data are presented in Figure Figure 1, showing that both round and flat brushes follow a power-law relation that depends solely on the tip shape, regardless of their intended use. Such a scaling reads:

$$L \sim D^{\beta}. \tag{1}$$

The exponent β is determined by the tip shape. Grouping the data according to the three main categories of tip geometries — rounded, conical and flattened ends — and fitting the relation (1) for each subset yields $\beta_{\mathrm{r}} \simeq 0.62 \pm 0.02$, $\beta_{\mathrm{c}} \simeq 0.54 \pm 0.04$ and $\beta_{\mathrm{f}} \simeq 0.90 \pm 0.04$. These exponents are robust across the different brush uses (fine art, nail art, construction) and materials in our dataset, and show that brushes are not always homothetic: their aspect ratio systematically changes with size in a tip-shape-dependent way. To understand the origin of these universal behaviors, it is therefore necessary to examine how brushes are used. After being impregnated with liquid, a brush is applied to a surface with a specific angle causing the bristles to buckle to varying degrees depending on the applied force. The brush is then slid across the surface to deposit the liquid it holds.

**Mechanical response**

The mechanical properties of the brushes are first characterized by analyzing their deformation under compressive loading in the absence of any liquid. The brushes are compressed against a flat surface considered infinitely rigid. The surface consisted of a 1.5 cm thick aluminum plate, covered with a sheet of conventional office paper (Inacopia Office 80 g) fixed with adhesive tape. The brush is mounted onto the load cell of an Adamel Lhomargy traction machine. We measured the force exerted by the brush on the sensor at a given applied angle as a function of brush indentation $z$. Measurements are taken during both the compression (loading) and decompression (unloading) stages.

In Figure 2($a$), a typical force $F$ evolution as a function of the compression $z$ is shown for three successive load/unload cycles performed on the same brush at angle $\theta = 0°$ (where $\theta$ is the angle between the paintbrush and the vertical axis as defined in Figure 2($c$)). No sign of aging is observed.

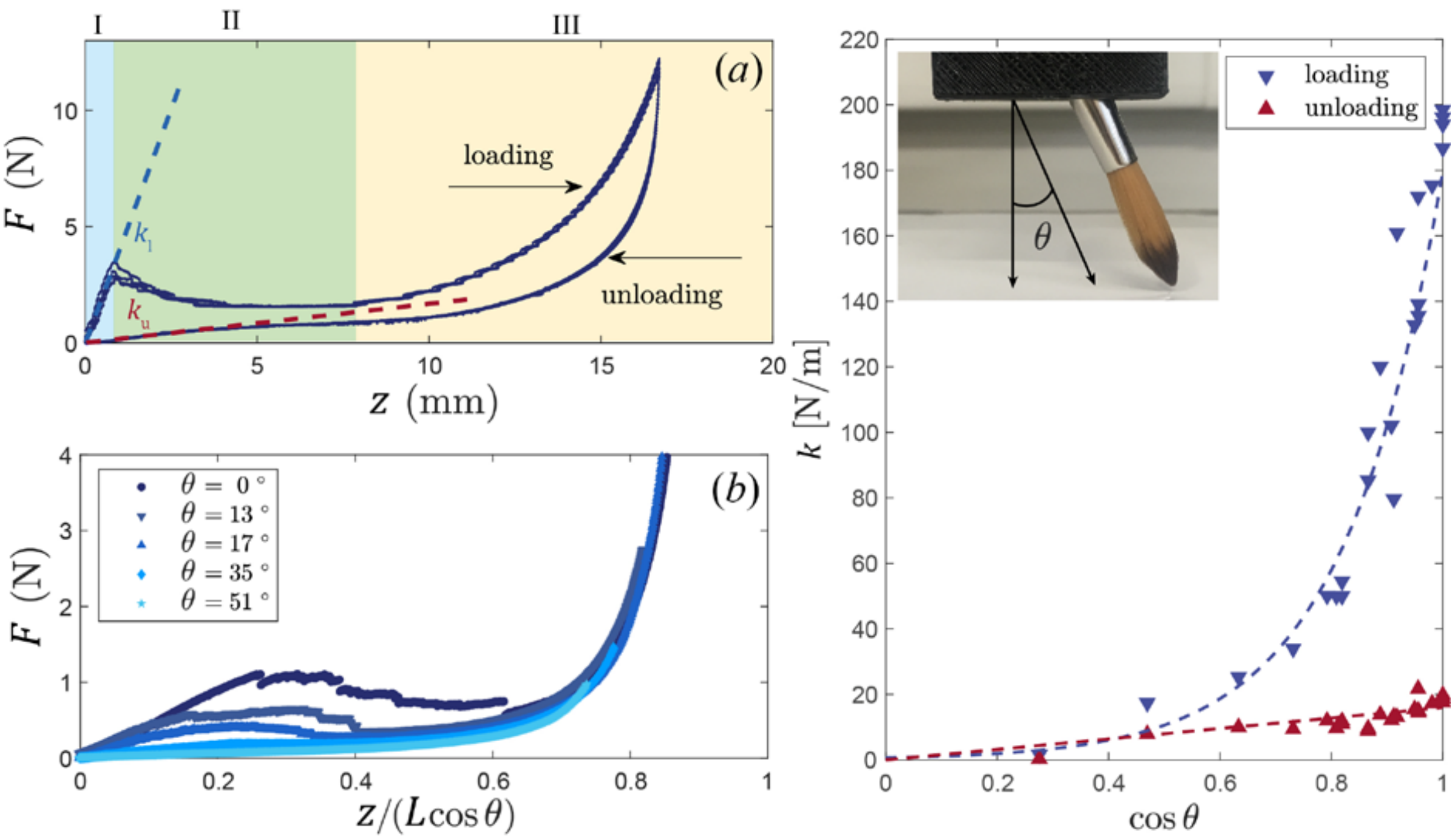

*Figure 2:* ***Force and stiffness as functions of brush indentation and application angle.*** *(a) Compression–decompression cycle for a HEMA Stippling brush 304 (L = 3 cm) on a Plexiglas surface, showing the successive phases of the process. The dashed lines indicate linear fits used to extract the loading and unloading stiffnesses. (b) Compression force as a function of rescaled indentation for a DALER-ROWNEY Dalon D77 Size 10 brush (L = 2.5 cm) at different application angles. (c) Loading and unloading stiffnesses as functions of the application angle for the same brush. Dashed lines indicate an exponential trend* $F = 0.63\,exp(5.66\cos\theta)$ *for the loading stiffness and a linear trend for the unloading stiffness, consistent with the normal projection of the applied force.*

The loading and unloading phases can each be divided into three distinct parts. This nonlinear force-displacement behavior is consistent with models of slender elastic structures under compression, where the initial linear regime (I) reflects the bending of individual fibers, followed by a plateau (II) dominated by friction, and a final nonlinear increase (III) as the structure approaches full compression characterized by $z \approx L$ (*23*). Similar multi-regime responses have been observed in brush seals and toothbrush bristles, where the collective mechanics of fiber bundles are governed by a competition between elastic deformation, friction, and geometric constraints (*24*, *25*). The hysteresis observed in our experiments further aligns with these systems, where the directionality of motion introduces frictional dissipation that depends on the roughness of the contact between fibers and the surface. During the loading phase, the force $F$ initially increases linearly with the displacement for small indentations, up to a peak force, corresponding to elastic deformation of the brush, and where we can define a stiffness $k$ during both compression $k_{\mathrm{l}}$ and decompression $k_{\mathrm{u}}$. The extent and amplitude of this initial linear regime strongly depend on the brush angle (see Figure 2(b)). Figure 2(c) reveals two distinct angular responses. Upon unloading, the stiffness follows the normal projection of a force applied along the brush axis with $k_{\mathrm{u}} \propto \cos(\theta)$. During loading, however, it increases more steeply than a cosine dependence, indicating an additional contribution from static friction and fiber bending.

At small deformation and for $\theta = 0°$, we measure the stiffness of all the tested brushes from the experimental curves $F(z)$. We observe that the stiffness remains almost independent of the length $L$, or the diameter $D$, both during compression ($k_{\mathrm{l}} \simeq 203 \pm 114$ N/m) and decompression phases ($k_{\mathrm{u}} \simeq 27 \pm 15$ N/m). To test this observation, in Figure 3, we compare the stiffness of brushes from a series of fine art brushes, whose diameters span over more than a decade from 1.1 mm to 25 mm, with stiffness measurements performed on one brush ($D = 10.5$ mm and $L = 38$ mm), but whose length is progressively reduced by cutting the fibers between each experiment. In this case, the stiffness strongly depend on their length. The length dependence of the stiffness is captured by the classical elastica response of independently bending fibers:

$$k = \frac{3NEI}{L^3}. \tag{2}$$

For the tested brush, this description gives $k = 250$ N/m, using an initial length $L = 38$ mm, $N \approx 3 \times 10^4$ fibers, a single-fiber Young's modulus $E \simeq 2$ GPa, and a second moment of area $I = \pi\rho^4/4$ for fibers of radius $\rho = 100\ \mu$m. Thus, the empirical length–diameter scaling of commercial brushes appears tuned to preserve a nearly size-independent stiffness.

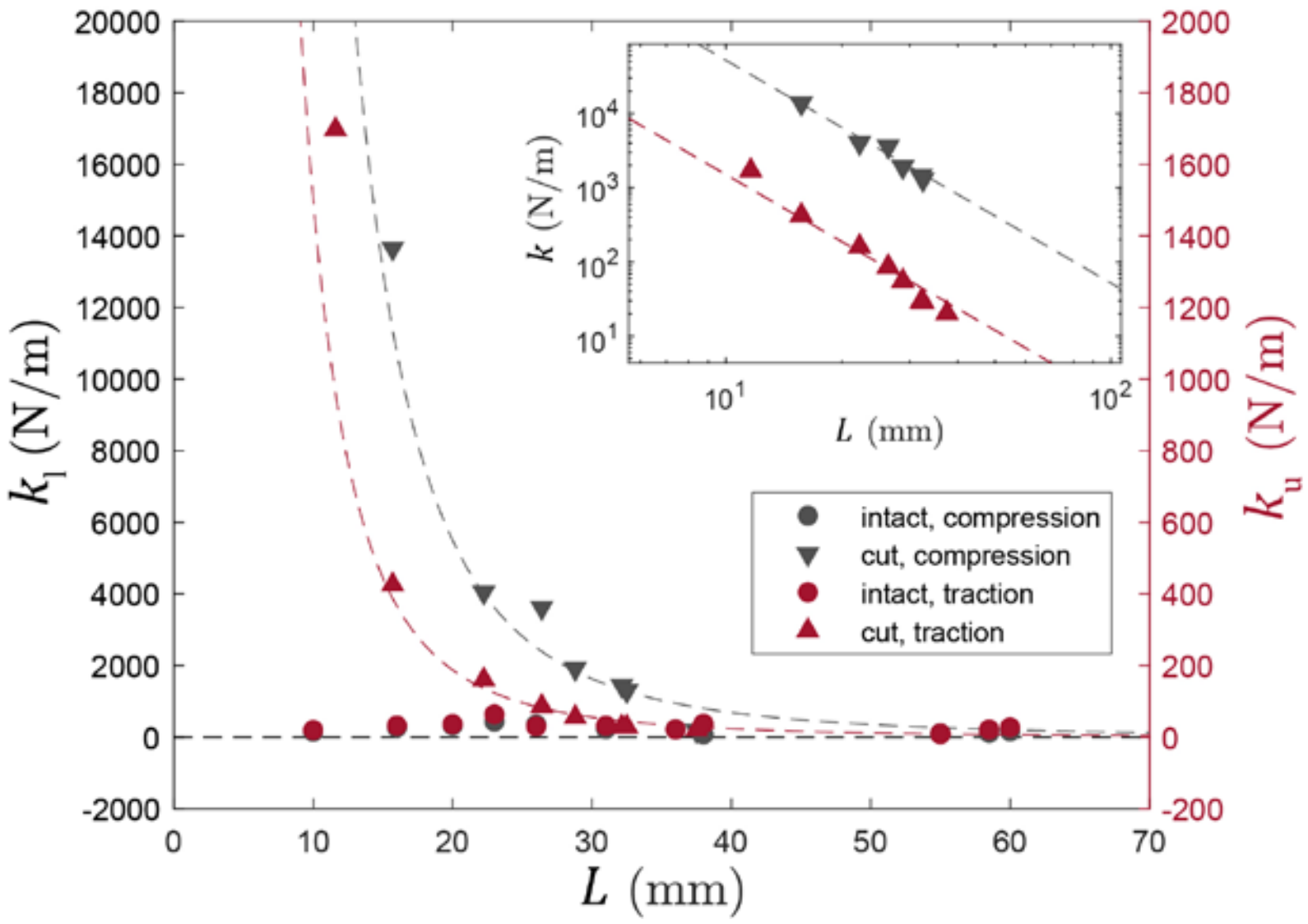

*Figure 3: **Stiffnesses $k_l$ and $k_u$, respectively for the loading and unloading phases, as a function of the brush length L**. Measurements have been done for DALER-ROWNEY Graduate Round brushes with different intact brushes of the same series and on a brush (Size 16, D=10 mm, L=3.8 cm), whose length is progressively shortened between each experiments. Data for "intact, compression" and "intact, traction" are superimposed in this representation.*

To better understand why a constant stiffness is chosen across different brushes, we develop a contact model that considers the geometry of the brush tip, the part that first enters into contact with the substrate under low stress (or strain) where the stiffness is measured. We distinguish three tip shapes: flat, rounded, and conical as seen on Figure 4. Each geometry leads to a different distribution of contact forces and deformation modes. In the flat-tip configuration, the contact is established over a finite area from the outset, resulting in an immediate collective response from multiple fibers. In the rounded-tip case, the contact initiates at a point and progressively spreads as the load increases, leading to a gradual increase in the apparent stiffness. For conical tips, the contact starts on a narrow region but involves bending of longer, less supported fibers. The tips shapes can be described using a power law of the form:

$$\frac{h}{L} \propto \left(\frac{r}{D}\right)^{\alpha}, \quad (3)$$

where $h$ (respectively $r$) the longitudinal (resp. transversal) distances to the tip, and $L$ (respectively $D$) the longitudinal (resp. transversal) brush sizes. The exponent $\alpha$ characterizes the tip geometry: $\alpha = 1$ corresponds to a conical tip, $\alpha = 2$ to a rounded (parabolic) tip, and $\alpha \rightarrow \infty$ to a flat tip with an abrupt edge.

When the brush is indented by a distance $\delta$, the fibers inside a circular section of surface $\pi r^2$, with $r \propto D \times (\delta/L)^{1/\alpha}$ are in contact with the solid surfaces. Let's $n$ the surface density of fibers, and:

$$F_s \propto \frac{EI}{L^2}, \quad (4)$$

the buckling force of one single fibers (*26*). The total force is then:

$$F \propto n\pi EI\delta^{2/\alpha}\left(L^{-2-2/\alpha}D^2\right). \quad (5)$$

This force does not depend on the brush sizes $D$ and $L$ if $L^{-2-2/\alpha}D^2$ is constant, so if $L \sim D^{\beta}$, with

$$\beta = \frac{\alpha}{\alpha + 1}, \quad (6)$$

which matches the experimental link between $L$ and $D$ for the three tested tip shapes described in Figure 1 ($\alpha = 1, \beta = 1/2 \simeq \beta_{\mathrm{c}}; \alpha = 2, \beta = 2/3 \simeq \beta_{\mathrm{r}}; \alpha = \infty, \beta = 1 \simeq \beta_{\mathrm{f}}$). This confirms that the observed scaling of the brush geometry leads to a size-independent effective stiffness, regardless of the brush function, size and tip shape.

**Impact on paint traces**

Such a size-independent stiffness is likely to impact the way the brush interacts with the surface, especially in terms of the shape of the paint traces. To investigate this, we measured the dimensions of the paint traces left by a series of round brushes and done by different paints at different concentrations. We varied the application angle and the imposed displacement as in Figure 4. First, we observed that the overall envelope of the force–displacement curves is not affected and the brush stiffness is the same with or without paint (see Figure S1). However, the small stick-slip events visible around the friction peak are completely smoothed due to the lubrication effect of the paint.

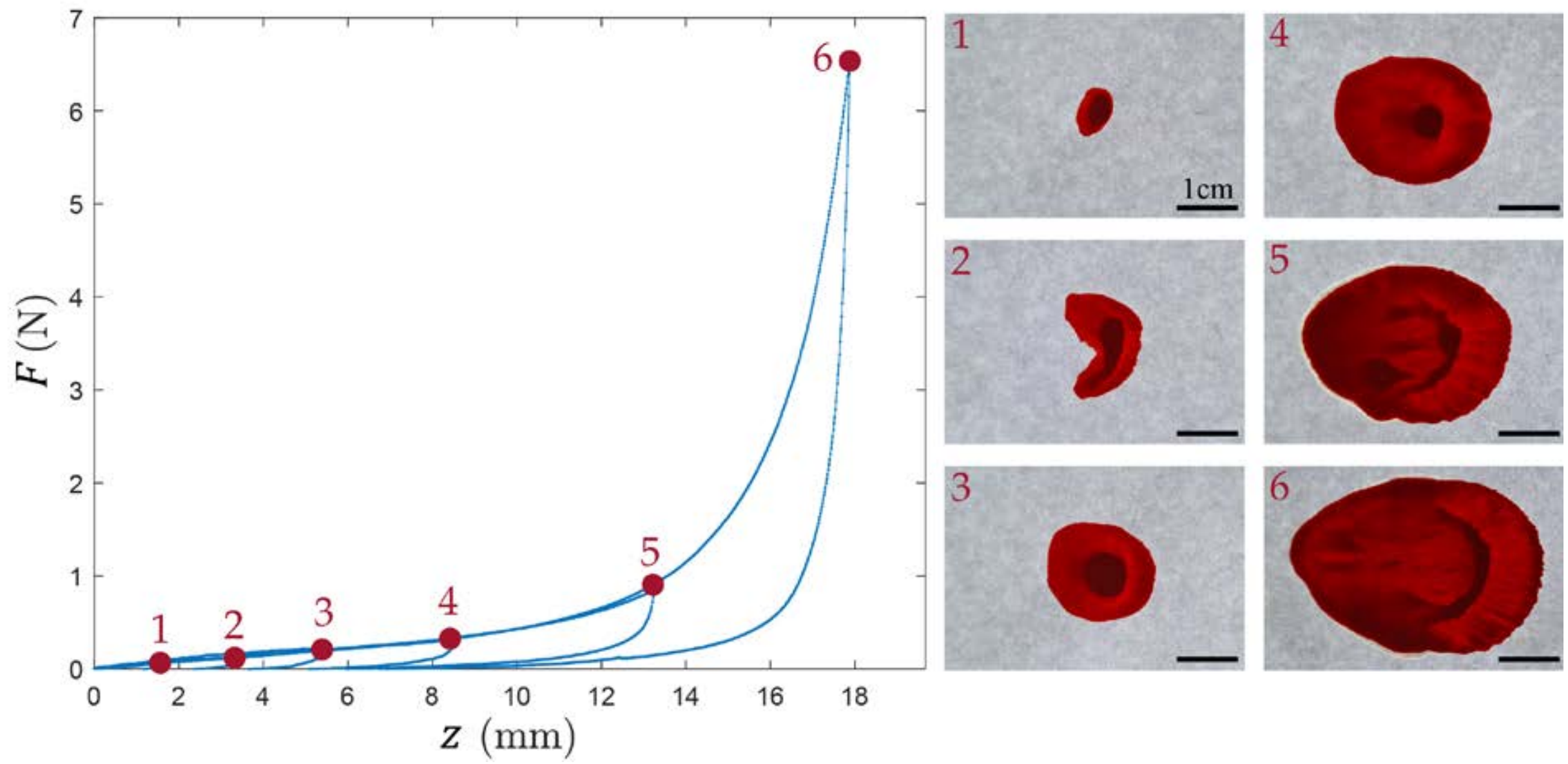

*Figure 4: **Force-displacement cycles associated with paint traces obtained at different maximal brush compression (numbered dots)** using a "Daler Rowney Graduate size 16" brush (rounded shape, rounded tip) applied at an angle $\theta = 43°$. The red paint is a FILA GIOTTO Extra Quality gouache, diluted in water with a 3:1 weight ratio (gouache to water). The support is a sheet of conventional office paper (Inacopia Office 80 g, A4 white).*

Despite these variations, we found that the surface area of the paint mark increases linearly with the imposed displacement.

To quantitatively test this relation, we analyzed the surface area of paint traces left by different round brushes as a function of the maximal brush compression in Figure 5. Remarkably, we observe that the surface area $A$ of the paint stain evolves linearly with the maximal brush compression $z_m$, regardless of the size of the brush used.

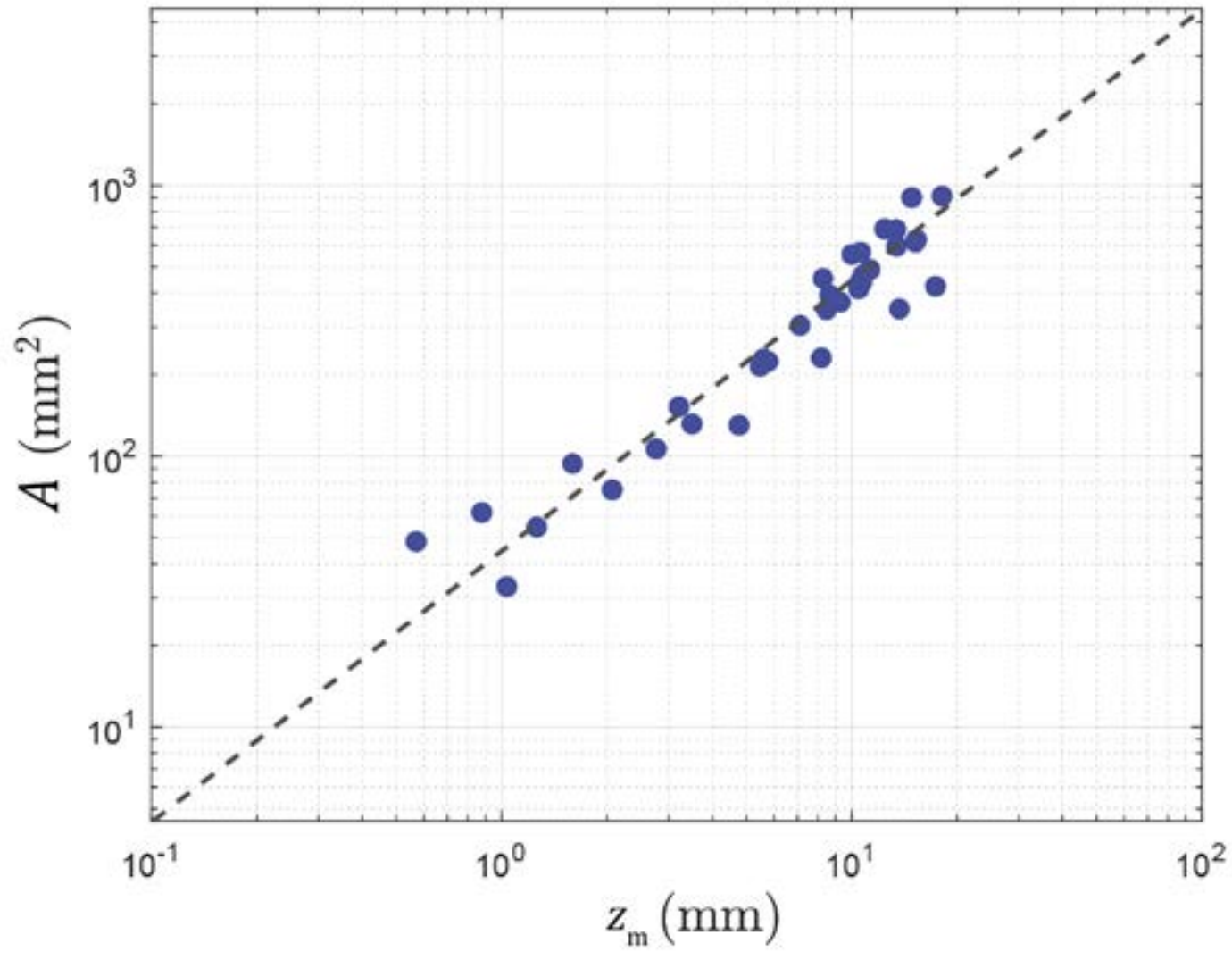

*Figure 5: **Surface area of paint traces as a function of maximal brush compression $z_m$ for different round brushes**. All measurements are performed using FILA GIOTTO Extra Quality gouache diluted in water (3:1 weight ratio) at an application angle α = 43°. The dashed line is a linear fit of the experimental data.*

By combining this evolution with the scaling law from Equation (1), we show that this independence stems directly from the constant stiffness of the brushes. Since $F = k.z_m$ and $k$ is almost constant for all brushes, the contact area, and therefore the size of the paint stain, depends only on the maximal brush compression $z_m$, which is itself proportional to the applied force. This relation has important implications for artistic practice: it first demonstrates that an artist, or a robot, can control trace sizes by only applying a given force, regardless of the rounded brush chosen.

**Conclusion**

Overall, these results show that the commercial geometry of the brush is not arbitrary: over time, it has converged over history toward shapes that maintain an essentially constant mechanical response over a wide range of sizes, rather than preserving a purely geometric proportion. This invariance is a direct result of the geometry of the tip controlling the relationship between the length of the bristles and the diameter of the ferrule. From this point of view, the constant stiffness measured on hundreds of brushes, covering two decades in diameter, used for different purposes and from several manufacturers, does not seem to be a coincidence, but the macroscopic signature of a fundamental geometric constraint imposed by brush users themselves.

This size-independent stiffness governs how round brushes deposit paint. Because the contact area scales directly with the applied compression rather than with the brush dimensions, an artist can predict and control the size of a paint stroke through force alone, independently of the specific round brush in hand. In this sense, paintbrushes join the set of everyday tools whose form has been progressively, perhaps implicitly, shaped by simple physical constraints rather than explicit design rules, in a Darwinian-like fashion for Human-made tools. The same approach, combining geometric measurements, mechanical testing, and a minimal contact model, could be extended to other slender, fiber-based applicators, from cosmetic brushes to industrial coating tools, to reveal further how generally such scaling principles hold.

**Acknowledgments:**

The authors thank the Soft Matter Collaborative Research Unit at Hokkaido University, Japan, and the International Research Network between France and India on “Hydrodynamics at Small Scales: From Soft Matter to Bioengineering”.

**Funding:**

The authors acknowledge financial support from the Agence Nationale de la Recherche through the Fricolas project (ANR-21-CE06-0039).

The authors also acknowledge financial support from the European Research Council through the EMetBrown grant (ERC-CoG-101039103) and the BIFFTANNEN grant (ERC-PoC-101247259). Views and opinions expressed are those of the authors only and do not necessarily reflect those of the European Union or the European Research Council.

**Author contributions:**

Conceptualization: JC, FR, CP

Investigation: RT, TS, CP

Visualization: RT, CP

Funding acquisition: FR, CP

Writing – original draft: CP

Writing – review & editing: RT, TS, JC, FR, CP

**Competing interests:**

Authors declare that they have no competing interests.

**Data and materials availability:**

All data are available in the main text.